# Alpha-tocopherol inhibits pore formation in the oxidized bilayers

Phansiri Boonnoy[1,2], Mikko Karttunen[3*], and Jirasak Wong-ekkabut[1,2*]

[1]Department of Physics, Faculty of Science, Kasetsart University, Bangkok, 10900, Thailand
[2]Computational Biomodelling Laboratory for Agricultural Science and Technology (CBLAST), Faculty of Science, Kasetsart University, Bangkok 10900, Thailand
[3]Department of Mathematics and Computer Science & Institute for Complex Molecular Systems, Eindhoven University of Technology, MetaForum, 5600 MB Eindhoven, the Netherlands

**ABSTRACT:** In biological membranes, alpha-tocopherols (α-toc; vitamin E) protect polyunsaturated lipids from free radicals. Although the interactions of α-toc with non-oxidized lipid bilayers have been studied, their on oxidized bilayers remain unknown. In this study, atomistic molecular dynamics (MD) simulations of oxidized lipid bilayers were performed with varying concentrations of α-toc. Bilayers with 1-palmitoyl-2-lauroyl-sn-glycero-3-phosphocholine (PLPC) lipids and its aldehyde derivatives at 1:1 ratio were studied. Our simulations show that oxidized lipids self-assemble into aggregates with a water pore rapidly developing across the lipid bilayer. The free energy of transporting an α-toc molecule in a lipid bilayer suggests that α-tocs can passively adsorb into the bilayer. When α-toc molecules were present at low concentrations in bilayers containing oxidized lipids, the formation of water pores was slowed down. At high α-toc concentrations, no pores were observed. Based on the simulations, we propose that the mechanism of how α-toc inhibits pore formation in bilayers with oxidized lipids is the following: α-tocs trap the polar groups of the oxidized lipids at the membrane-water interface resulting in a decreased probability for the oxidized lipids to reach contact with the two leaflets and initiate pore formation. This demonstrates that α-toc molecules not only protect the bilayer from oxidation but also help to stabilize the bilayer after lipid peroxidation occurs. These results will help in designing more efficient molecules to protect membranes from oxidative stress.

Biological membranes serve as a partition between cells and their environment. Under oxidative stress, unsaturated lipids present in cell membranes may become exposed to attacks by free radicals, that is, oxidation. Oxidation transforms some of the membrane lipids to oxidized ones such as hydroperoxide and aldehyde lipids.[1,2] It has also been suggested that internal, that is intra-leaflet, oxidation may be important in altering bilayer properties.[3]

Lipid peroxidation is an important mechanism of cell membrane damage.[4-6] Previous experiments and computer simulations[4,7-14] have demonstrated how oxidized lipids disturb and deform bilayers. The polar chains in oxidized lipids are energetically unfavorable to stay in the bilayer's interior resulting in the reversal of the polar lipid chain to the bilayer interface.[4,11,15,16] This reversal causes major changes in bilayer properties such as increase of area per lipid, bilayer thinning, decrease of lipid tail order parameter, and increase in water permeability.[4,12,15-18] Recently, we performed MD simulations of lipid bilayers with oxidized lipids at high concentrations. Two major oxidized lipid species including hydroperoxide and aldehyde were studied. The results showed that only aldehyde lipids were able to induce pore formation across a PLPC lipid bilayer and cause significant bilayer deformation.[13,15]

α-toc is well-known as an efficient antioxidant that protects membranes from free radical-initiated oxidation.[19-21] Natural membranes consist of saturated and unsaturated lipids, and they are permeable to water and small molecules. Unsaturated lipids play an important role in membrane permeability by disrupting the packing of saturated lipids. However, unsaturated lipids are readily susceptible to peroxidation which, if extreme, may lead to uncontrollable transport of molecules across the membrane.[22] Numerous studies have been conducted to find the mechanisms of how α-toc protects polyunsaturated fatty acids from free radicals and to explain how α-toc interacts with biological membranes.[21,23-28] Protective mechanisms in the absence of oxidized lipids have been proposed, for example that the chromanol group of the α-toc molecule can bind and trap free radicals within the interior and near the membrane interface[26] thus blocking their ability to enter the membrane and oxidize polyunsaturated lipid chains.

In previous studies, interactions of α-toc have been considered with non-oxidized lipid bilayers only[24-28] and the effects of α-toc on bilayers with oxidized lipids remain unresolved. To understand how α-tocs interact with biological membranes after lipid peroxidation, MD simulations of α-toc in oxidized lipid bilayers were carried out. Previous studies[13,15] without α-toc have shown that the presence of aldehyde lipids could lead to large disturbances of the bilayer. Hence, the 1:1 binary lipid bilayer mixture between 1-palmitoyl-2-linoleoyl-sn-glycero-3-phosphatidylcholine (PLPC) and its two aldehyde lipids 1-palmitoyl-2-(9-oxo-nonanoyl)-sn-glycero-3-phosphocholine (9-al) and 1-stearoyl-2-(12-oxo-cis-9-dodecenoyl)-snglycero-3-phosphocholine (12-al) lipids were chosen for this study.

**Methods:** The simulated systems consisted of 1-16 α-toc molecules embedded in lipid bilayers with 128 phospholipid molecules and 10,628 simple point charge (SPC)[29] water molecules. All the simulated systems are listed in Table 1. The topologies and force field parameters of α-toc were taken from Qin et al.[24,25] The lipid parameters were taken from previous studies.[4,13,30] Initially, the α-toc molecules were randomly placed in the water phase about 4.2 nm away from the center of mass (COM) of the bilayer. After energy minimization, MD simulations were run for 1-2 μs with 2 fs integration time step by using the GROMACS 4.5.5 package.[31,32] Periodic boundary conditions were applied in all directions. The neighbor list was updated at every time step. A cutoff was employed at 1.0 nm for the real space part of electrostatic interactions and Lennard-Jones interactions. The particle–mesh Ewald[33-35] was used to calculate the long-range part of electrostatic interactions and all bond lengths were constrained by the LINCS algorithm.[36] In all simulations, the temperature was set to 298 K using the v-rescale algorithm[37] with a time constant of 0.1 ps. Pressure was controlled by the Parrinello-Rahman algorithm[38,39] with an equilibrium semi-isotropic pressure of 1 bar, a time constant of 4.0 ps and compressibility of $4.5\times10^{-5}$ bar$^{-1}$. These parameters and protocols have been extensively tested and optimized.[40-42] All visualizations were done using Visual Molecular Dynamic (VMD) software.[43]

**Free energy calculation:** The potential of mean force (PMF) of an α-toc transferring into the lipid bilayer was calculated using the umbrella sampling technique[44] with the Weighted Histogram Analysis Method[45] (WHAM). Three bilayers of 100% PLPC, 50% 12-al, and 50% 9-al were used. A series of 41 simulations was run with the distance between α-toc and the bilayer center restrained between 0 and 4.0 nm, with 0.1 nm spacing. In the first window, the α-toc molecule is in bulk water. It was then subsequently moved into the bilayer along the bilayer normal (z-axis) in each successive window. Therefore, the final window had the α-toc molecule at the bilayer center. The hydroxyl of α-toc was restrained with respect to the center of mass of the bilayer, using a harmonic restraint with a force constant of 3000 kJ/(mol nm$^2$) normal to the bilayer. Simulations were performed in the NPT ensemble at 298 K for total time of 2.05 μs (50 ns per each window). The bootstrap analysis method[46] was used to estimate the statistical uncertainty in umbrella sampling simulations.

**Results:** The MD simulations show that without α-toc, oxidized lipids self-assemble to form aggregates and that a water pore develops rapidly across the bilayer (Figure 1). A pore spanning the bilayer occurred after 140 and 180 ns in 50% 12-al and 9-al bilayers, respectively. This result is in agreement with previous studies[13]. At low α-toc concentrations (2 and 4 α-toc molecules in a bilayer), α-toc molecules' preferred position was close to the bilayer interface which lead to a slowed down formation (as compared with no α-toc present) of a water pore over several hundreds of nanoseconds, Table 1. Interestingly, when α-toc concentration was increased, no pore formation was observed over the entire simulation time of over 2 μs.

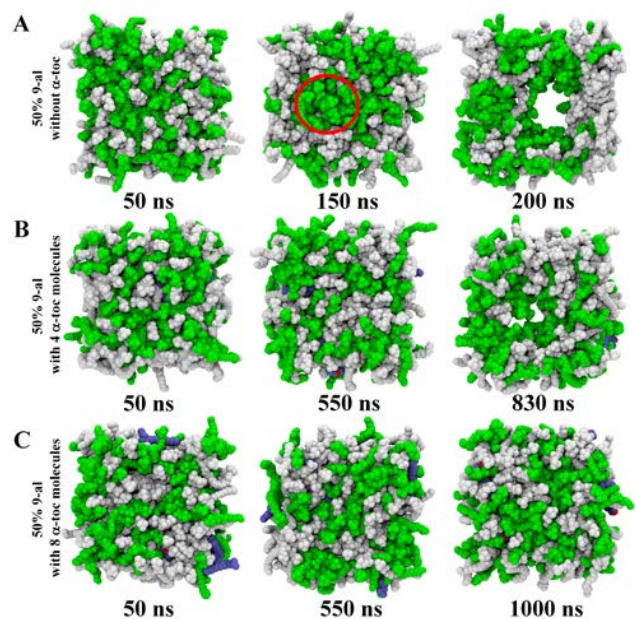

**Figure 1.** Time evolution of a 50% 9-al lipid bilayer A) without α-toc, B) with 4 α-toc, and C) 8 α-toc molecules. Water molecules are not shown for clarity. White, green, purple: PLPC, 9-al and α-toc molecules, respectively. Red: Oxygen atoms of the α-toc hydroxyl groups. The red circle represents the oxidized lipids' aggregation region.

Figure 2 shows (unbiased MD simulation) that α-toc is able to passively penetrate into the bilayer and remain in the bilayer interior around the carbonyl groups of the lipids. In agreement with free energy calculations (Figure 3), it is energetically favorable for α-toc to enter the bilayer. The free energy calculation shows that the adsorption energy of α-toc into a PLPC bilayer is -43.2 kJ/mol with the equilibrium position being about 1.2 nm from the center of bilayer. For moving the α-toc from equilibrium toward the center of the bilayer, an energy barrier with a steep slope and magnitude of 12.7 kJ/mol was observed

(Figure 3). After reaching the maximum of the free energy barrier, the PMF plateaus at -30.5 kJ/mol. Our PMF profile is qualitatively similar to that of a cholesterol molecule transferring from water into a DPPC lipid bilayer.[47] As a comparison, the free energies of cholesterol transferring from water in DPPC lipid bilayers to the center of bilayer and equilibrium are -50 and -67 kJ/mol, respectively.[47] This result suggests that cholesterol is more favorable to stay in lipid bilayer than an α-toc molecule. Note that some of the quantitative differences might come from other parameters such as type of lipid and temperature, and thus further studies are needed to establish the differences more accurately.[47]

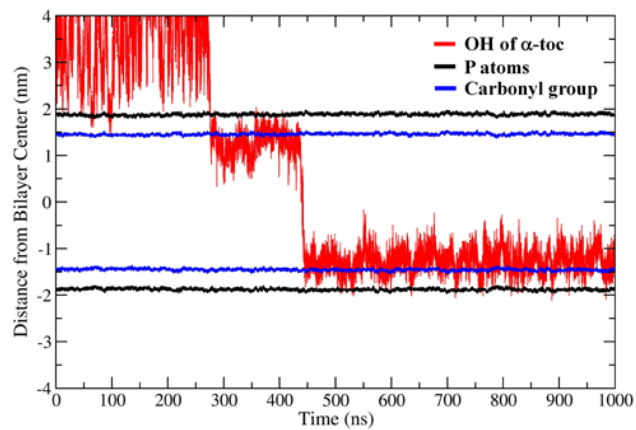

**Figure 2.** The time evolution of the position of the hydroxyl group of an α-toc molecule in 100% PLPC bilayer.

The free energy barrier of an α-toc between equilibrium and bilayer's center was 12.7 kJ/mol (Figure 3) and as a result, flip-flop of an α-toc in the pure PLPC bilayer was observed only once. The PMF profiles of α-toc are qualitatively similar to non-oxidized bilayer but the adsorption energies were increased to -50.8 and -49.7 kJ/mol in 50% 12-al and 50% 9-al bilayers, respectively. The equilibrium positions of α-tocs were deeply inside bilayer consistent with a decrease in the thickness of the bilayer. These results suggest that α-toc molecules prefer to interact with oxidized lipids and stay inside the oxidized bilayer as compared to the non-oxidized system. Moreover, the free energy barriers from equilibrium toward the center of the bilayer of 50% 12-al and 50 % 9-al were 5.9 and 7.1 kJ/mol, respectively. The observed decreases of the free energy barriers in oxidized bilayers results in frequent flip-flops of α-toc (Figure 4). Moreover, α-tocs always formed hydrogen bonds with the aldehyde groups oxidized lipids' tails. Flip-flops of α-tocs occurred when such hydrogen bonds were lost and re-formed with aldehyde groups in the opposite leaflet as shown in Figures 4 and S1. In membranes containing no oxidized lipids, α-toc flip-flops have previously been observed only at the temperature of 350K but not below.[24] These results suggest that α-toc is highly mobile inside bilayers containing oxidized lipids.

Electron density distributions (Figure 5) show that the hydroxyl groups of α-toc molecules have their maxima at 0.96 nm and 1.06 nm for the 12-al and 9-al systems, respectively. These maxima are related to the locations of the carbonyl groups of the lipid bilayers and consistent with Figure 4 and previous studies.[24,26-28] Surprisingly, the electron density of the oxygen atoms in oxidized lipids' tails decreased at the bilayer's center and increased at the water interface when α-toc molecules were present. Our previous study suggested that one of the key mechanisms for passive pore formation is the distribution of polar groups inside the bilayer.[13] The deep penetration of the polar group inside bilayer can bring lipids in contact with lipids in the opposite leaflet thus leading to the formation of a water bridge and consequently a stable pore.

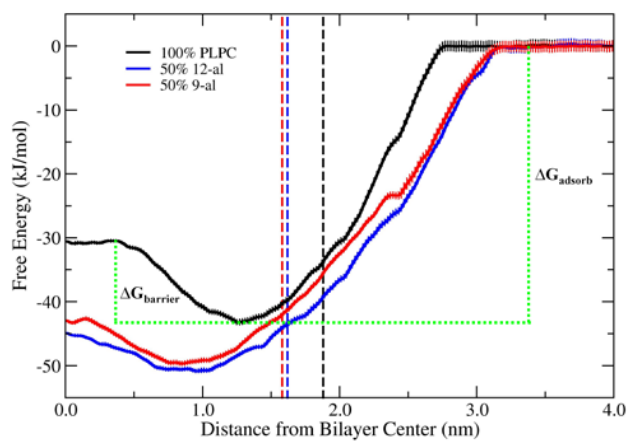

**Figure 3.** The potential of mean force (PMF) of an α-toc transferring into the bilayer as a function of distance in the z-direction from the center of bilayer (z = 0.0 nm). The curves show bilayers of 100% PLPC, 50% 12-al and 50% 9-al in black, blue, and red lines, respectively. Bulk water defines zero free energy. The dashed lines represent the average position of P atoms in each bilayer. The free energies of α-toc adsorbing in the 100% PLPC, 50% 12-al, and 50% 9-al bilayer are -43.2, -50.8, and -49.7 kJ/mol, respectively. The free energies barriers of α-toc in the 100% PLPC, 50% 12-al, and 50% 9-al bilayers are 12.7, 5.9, and 7.1 kJ/mol, respectively. The maximum statistical uncertainties of 100% PLPC, 50% 12-al, and 50% 9-al are 0.8, 0.7, and 0.9 kJ/mol, respectively.

When α-tocs are present in a bilayer, they are able to trap the polar groups of the oxidized lipids at the water interface (Figures 4 and 5) resulting in a decreased probability for the oxidized lipids to be in contact with each other. Therefore, pores cannot be formed at high α-toc concentrations. Furthermore, water transport across lipid bilayers with non-oxidized lipids is not frequent, since the free energy barrier of a water molecule crossing a PLPC bilayer is 29.4±2.3 kJ/mol.[4] Permeability also increases by an order of magnitude as the concentration of oxidized lipids increases.[17,18] Moreover, permeability increases by two orders of magnitude when a pore spans the bilayer (Table 2). When the concentration of α-toc inside the bilayer increases, water permeability decreases and pore formation becomes inhibited (Table 2).

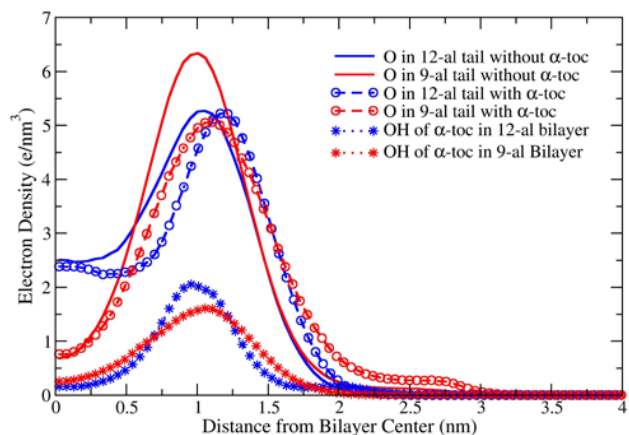

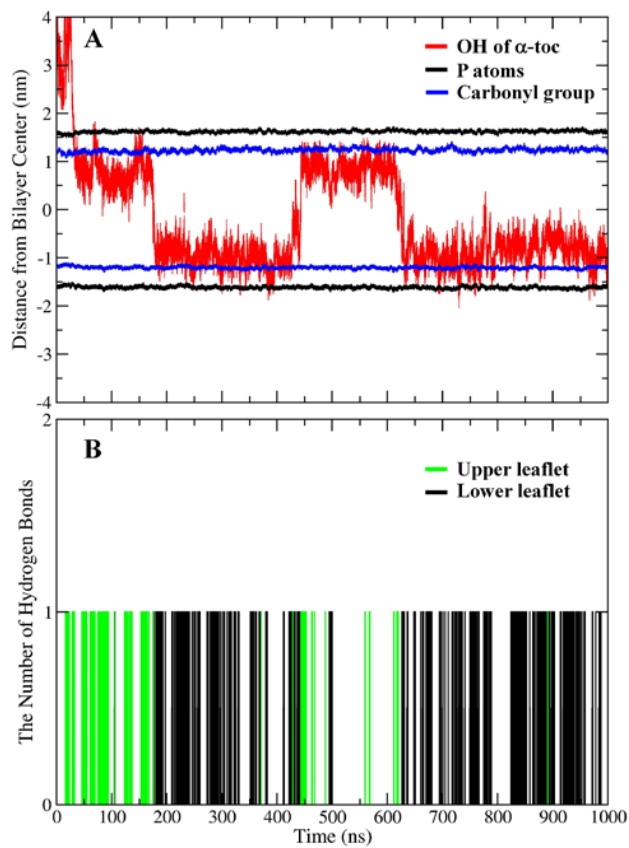

**Figure 4.** A) The time evolution of the position of the hydroxyl group of α-toc from the center of 50% 9-al lipid bilayer (along the z-axis). B) The number of hydrogen bonds between the hydroxyl groups of α-tocs and the aldehyde groups of the oxidized lipid tails in the upper and lower leaflets. This result is from only one of eight α-toc molecules in system of 50% 9-al with 8 α-toc and the rest of α-toc molecules are shown in Figure S1 (Supporting Information)

Cholesterol and α-toc molecules have similar molecular structures consisting of a hydrophobic tail and a ring-structure with a hydroxyl group. The effects these molecules on biological membranes are similar and that has been suggested for being responsible for the observed increase of bilayer thickness and lipid tail order.[25,48,49] Moreover, Issack et al.[50] have shown that the free-energy barrier for transferring a water molecule to the center of the bilayer was increased by 6 kJ/mol when 41 mol% cholesterol was present in a DPPC bilayer. This observed significant increase of the free energy barrier to transfer water through bilayer with the presence of cholesterol results in a decrease in water permeability.[50,51]

**Figure 5.** Electron density profiles for the oxygen atoms in the oxidized lipid hydrocarbon chains with and without α-toc, and for the hydroxyl groups of α-toc in 50% 12-al and 9-al bilayers. All bilayers consisted of 8 α-toc molecules.

In conclusion, free radicals play an important role in membrane damage and aldehyde lipids are the major oxidative lipid product that causes pore formation and bilayer deformation.[13,19-21] On the other hand, α-toc is one of most effective antioxidants in removing free radicals and it has been used in cosmetics, functional foods and many other applications.[52-54] Previously[19-21,26], the only protective mechanism of α-toc to against lipid peroxidation in biological membranes was proposed to be due to α-toc blocking free radicals' entry into the membrane thus protecting polyunsaturated lipid chains from oxidation process, that is, with no oxidized lipids in the bilayer. In a realistic case, however, oxidized lipids are present and it is important to understand how their destructive effects can be prevented. In this study, we have shown that α-toc molecules can inhibit pore formation in oxidized lipid bilayers by confining the polar groups of the oxidized lipids at the water interface. Our findings also suggest that by controlling α-toc concentration, the stability of biological membranes can be increased. This understanding of how α-tocs affect oxidized membranes is likely to be beneficial for designing new molecules to protect, e.g., skin, against aging[55,56] and in plasma treatment of cancer[48].

**AUTHOR INFORMATION**
**Corresponding Authors**
*J.W.: jirasak.w@ku.ac.th
**Notes**
The authors declare no competing financial interest.

**ACKNOWLEDGMENTS**
Financial support by Kasetsart University Research & Development Institute (KURDI; J.W.) and Faculty of Science (J.W.) at Kasetsart University.

**REFERENCES**

(1) Mandal, T. K.; Chatterjee, S. N. *Radiat. Res.* **1980**, *83*, 290-302.
(2) Chatterjee, S. N.; Agarwal, S. *Free Radic. Biol. Med.* **1988**, *4*, 51-72.
(3) Else, P. L.; Kraffe, E. *Biochim. Biophys. Acta* **2015**, *1848*, 417-421.
(4) Wong-ekkabut, J.; Xu, Z.; Triampo, W.; Tang, I. M.; Peter Tieleman, D.; Monticelli, L. *Biophys. J.* **2007**, *93*, 4225-4236.
(5) Yin, H. Y.; Xu, L. B.; Porter, N. A. *Chem. Rev.* **2011**, *111*, 5944-5972.
(6) Niki, E.; Yoshida, Y.; Saito, Y.; Noguchi, N. *Biochem. Biophys. Res. Commun.* **2005**, *338*, 668-676.
(7) Li, X. M.; Salomon, R. G.; Qin, J.; Hazen, S. L. *Biochemistry* **2007**, *46*, 5009-5017.
(8) Mattila, J. P.; Sabatini, K.; Kinnunen, P. K. *Langmuir* **2008**, *24*, 4157-4160.
(9) Sabatini, K.; Mattila, J. P.; Megli, F. M.; Kinnunen, P. K. J. *Biophys. J.* **2006** *90*, 4488-4499.
(10) Beranova, L.; Cwiklik, L.; Jurkiewicz, P.; Hof, M.; Jungwirth, P. *Langmuir* **2010**, *26*, 6140-6144.
(11) Khandelia, H.; Mouritsen, O. G. *Biophys. J.* **2009**, *96*, 2734-2743.
(12) Jarerattanachat, V.; Karttunen, M.; Wong-ekkabut, J. *J. Phys. Chem. B* **2013**, *117*, 8490-8501.
(13) Boonnoy, P.; Jarerattanachat, V.; Karttunen, M.; Wong-ekkabut, J. *J. Phys. Chem. Lett.* **2015**, *6*, 4884-4888.
(14) Siani, P.; de Souza, R. M.; Dias, L. G.; Itri, R.; Khandelia, H. *Biochim. Biophys. Acta* **2016**, *1858*, 2498-2511.
(15) Cwiklik, L.; Jungwirth, P. *Chem. Phys. Lett.* **2010**, *486*, 99-103.
(16) Garrec, J.; Monari, A.; Assfed, X.; Mir, L. M.; Tarek, M. *J. Phys. Chem. Lett.* **2014**, *5*, 1653-1658.
(17) Lis, M.; Wizert, A.; Przybylo, M.; Langner, M.; Swiatek, J.; Jungwirth, P.; Cwiklik, L. *Phys. Chem. Chem. Phys.* **2011**, *13*, 17555-17563.
(18) Runas, K. A.; Malmstadt, N. *Soft Matter* **2015**, *11*, 499-505.
(19) Niki, E.; Traber, M. G. *Ann. Nutr. Metab.* **2012**, *61*, 207-212.
(20) Wang, X.; Quinn, P. J. *Prog. Lipid Res.* **1999**, *38*, 309-336.
(21) Traber, M. G.; Atkinson, J. *Free Radic. Biol. Med.* **2007**, *43*, 4-15.
(22) Brodnitz, M. H. *J. Agric. Food Chem.* **1968**, *16*, 994-999.
(23) Kagan, V. E. *Ann. N. Y. Acad. Sci.* **1989**, *570*, 121-135.
(24) Qin, S. S.; Yu, Z. W.; Yu, Y. X. *J. Phys. Chem. B* **2009**, *113*, 16537-16546.
(25) Qin, S. S.; Yu, Z. W. *Acta Phys-Chim. Sin.* **2011**, *27*, 213-227.
(26) Leng, X. L.; Kinnun, J. J.; Marquardt, D.; Ghefli, M.; Kucerka, N.; Katsaras, J.; Atkinson, J.; Harroun, T. A.; Feller, S. E.; Wassall, S. R. *Biophys. J.* **2015**, *109*, 1608-1618.
(27) Marquardt, D.; Kučerka, N.; Katsaras, J.; Harroun, T. A. *Langmuir* **2014**, *31*, 4464-4472.
(28) Marquardt, D.; Williams, J. A.; Kucerka, N.; Atkinson, J.; Wassall, S. R.; Katsaras, J.; Harroun, T. A. *J. Am. Chem. Soc.* **2013**, *135*, 7523-7533.
(29) Berendsen, H. J. C.; Postma, J. P. M.; van Gunsteren, W. F.; Hermans, J. In *Intermolecular Forces*; Pullman, B., Ed.; D. Reidel: Dordrecht, The Netherlands, 1981, p 331-342.
(30) Bachar, M.; Brunelle, P.; Tieleman, D. P.; Rauk, A. *J. Phys. Chem. B* **2004**, *108*, 7170-7179.
(31) Hess, B.; Kutzner, C.; van der Spoel, D.; Lindahl, E. *J. Chem. Theory Comput.* **2008**, *4*, 435-447.
(32) Pronk, S.; Pall, S.; Schulz, R.; Larsson, P.; Bjelkmar, P.; Apostolov, R.; Shirts, M. R.; Smith, J. C.; Kasson, P. M.; van der Spoel, D.; Hess, B.; Lindahl, E. *Bioinformatics* **2013**, *29*, 845.
(33) Darden, T.; York, D.; Pedersen, L. *J. Chem. Phys.* **1993**, *98*, 10089-10092.
(34) Essmann, U.; Perera, L.; Berkowitz, M. L.; Darden, T.; Lee, H.; Pedersen, L. G. *J. Chem. Phys.* **1995**, *103*, 8577-8593.
(35) Karttunen, M.; Rottler, J.; Vattulainen, I.; Sagui, C. *Curr. Top. Membr.* 2008, 60, 49.
(36) Hess, B.; Bekker, H.; Berendsen, H. J. C.; Fraaije, J. G. E. M. *J. Comput. Chem.* **1997**, *18*, 1463-1472.
(37) Bussi, G.; Donadio, D.; Parrinello, M. *J. Chem. Phys.* **2007**, *126*, 014101.
(38) Parrinello, M.; Rahman, A. *J. Appl. Phys.* **1981**, *52*, 7182-7190.
(39) Nosé, S.; Klein, M. L. *Mol. Phys.* **1983**, *50*, 1055-1076
(40) Wong-Ekkabut, J.; Miettinen, M. S.; Dias, C.; Karttunen, M. *Nat. Nanotechnol.* **2010**, *5*, 555-557.
(41) Wong-ekkabut, J.; Karttunen, M. *J. Chem. Theory Comput.* **2012**, *8*, 2905-2911.
(42) Wong-ekkabut, J.; Karttunen, M. *Biochim. Biophys. Acta* **2016**, *1858*, 2529.
(43) Humphrey, W.; Dalke, A.; Schulten, K. *J. Mol. Graphics* **1996**, *14*, 33-38.
(44) Torrie, G. M.; Valleau, J. P. *J. Comput. Phys.* **1977**, *23*, 187-199.
(45) Kumar, S.; Rosenberg, J. M.; Bouzida, D.; Swendsen, R. H.; Kollman, P. A.; Rosenbergl, J. M. *J. Comput. Chem.* **1992**, *13*, 1011-1021.
(46) Hub, J. S.; Winkler, F. K.; Merrick, M.; de Groot, B. L. *J. Am. Chem. Soc.* **2010**, *132*, 13251-13263.
(47) Bennett, W. F. D.; MacCallum, J. L.; Hinner, M. J.; Marrink, S. J.; Tieleman, D. P. *J. Am. Chem. Soc.* **2009**, *131*, 12714-12720.
(48) Van der Paal, J.; Neyts, E. C.; Verlackt, C. C. W.; Bogaerts, A. *Chem. Sci.* **2016**, *7*, 489-498.
(49) Hofsass, C.; Lindahl, E.; Edholm, O. *Biophys. J.* **2003**, *84*, 2192-2206.

(50) Issack, B. B.; Peslherbe, G. H. *J. Phys. Chem. B* **2015**, *119*, 9391-9400.
(51) Saito, H.; Shinoda, W. *J. Phys. Chem. B* **2011**, *115*, 15241-15250.
(52) Tucker, J. M.; Townsend, D. M. *Biomed. Pharmacother.* **2005**, *59*, 380-387.
(53) Azzi, A.; Gysin, R.; Kempna, P.; Ricciarelli, R.; Villacorta, L.; Visarius, T.; Zingg, J. M. *Mol. Aspects Med.* **2003**, *24*, 325-336.
(54) Institute of Medicine, *"Vitamin E". Dietary Reference Intakes for Vitamin C, Vitamin E, Selenium, and Carotenoids*; The National Academies Press: Washington, DC, 2000.
(55) Lin, F. H.; Lin, J. Y.; Gupta, R. D.; Tournas, J. A.; Burch, J. A.; Selim, M. A.; Monteiro-Riviere, N. A.; Grichnik, J. M.; Zielinski, J.; Pinnell, S. R. *J. Invest. Dermatol.* **2005**, *125*, 826-832.
(56) Krutmann, J. *Hautarzt* **2011**, *62*, 576.

**Table 1.** Compositions of α-toc in the different oxidized lipid bilayer used in this study.

| No. | system name | proportion | Final structure | Pore time |
|---|---|---|---|---|
| 1 | 100% PLPC | 128 PLPC | Bilayer | - |
| 2 | 100% PLPC + 1 α-toc | 128 PLPC:1 α-toc | Bilayer | - |
| 3 | 50% 12-al | 64 PLPC: 64 12-al | Bilayer with a pore | 140 ns |
| 4 | 50% 12-al + 2 α-toc | 64 PLPC:64 12-al:2 α-toc | Bilayer with a pore | 337 ns |
| 5 | 50% 12-al + 4 α-toc | 64 PLPC:64 12-al:4 α-toc | Bilayer with a pore | 340 ns |
| 6 | 50% 12-al + 8 α-toc | 64 PLPC:64 12-al:8 α-toc | Bilayer | - |
| 7 | 50% 12-al + 16 α-toc | 64 PLPC:64 12-al:16 α-toc | Bilayer | - |
| 8 | 50% 9-al | 64 PLPC: 64 9-al | Bilayer with a pore | 180 ns |
| 9 | 50% 9-al + 2 α-toc | 64 PLPC:64 9-al:2 α-toc | Bilayer with a pore | 212 ns |
| 10 | 50% 9-al + 4 α-toc | 64 PLPC:64 9-al:4 α-toc | Bilayer with a pore | 809 ns |
| 11 | 50% 9-al + 8 α-toc | 64 PLPC:64 9-al:8 α-toc | Bilayer | - |
| 12 | 50% 12-al + 16 α-toc | 64 PLPC:64 9-al:16 α-toc | Bilayer | - |

**Table 2.** Water permeability through oxidized lipid bilayer

| No. | system name | Bilayer permeability ($p_f$) cm³/s (x $10^{-15}$) | Pore permeability ($p_f$) cm³/s (x $10^{-13}$) |
|---|---|---|---|
| 1 | 100% PLPC | 0.96 | - |
| 2 | 100% PLPC + 1 α-toc | 0.62 | - |
| 3 | 50% 12-al | 6.00 | 7.31 |
| 4 | 50% 12-al + 2 α-toc | 3.27 | 4.72 |
| 5 | 50% 12-al + 4 α-toc | 3.61 | 5.89 |
| 6 | 50% 12-al + 8 α-toc | 4.22 | - |
| 7 | 50% 12-al + 16 α-toc | 2.81 | - |
| 8 | 50% 9-al | 4.50 | 8.97 |
| 9 | 50% 9-al + 2 α-toc | 3.72 | 8.28 |
| 10 | 50% 9-al + 4 α-toc | 3.55 | 6.06 |
| 11 | 50% 9-al + 8 α-toc | 2.21 | - |
| 12 | 50% 12-al + 16 α-toc | 2.57 | - |

Note: Water permeability was calculated by $p_f = v_w R_t / NA$, where $v_w$ is the average volume of a single water molecule, 18 cm³/mol, $R_t$ is the rate of water transport across the bilayer and NA is Avogadro's number.

**Supporting Information**

**Alpha-tocopherol inhibits pore formation in the oxidized bilayers**

Phansiri Boonnoy[1,2], Mikko Karttunen[3], and Jirasak Wong-ekkabut[1,2*]

[1]Department of Physics, Faculty of Science, Kasetsart University, Bangkok, 10900, Thailand
[2]Computational Biomodelling Laboratory for Agricultural Science and Technology (CBLAST), Faculty of Science, Kasetsart University, Bangkok 10900, Thailand

[3]Department of Mathematics and Computer Science & Institute for Complex Molecular Systems, Eindhoven University of Technology, MetaForum, 5600 MB Eindhoven, the Netherlands

*Corresponding authors: jirasak.w@ku.ac.th

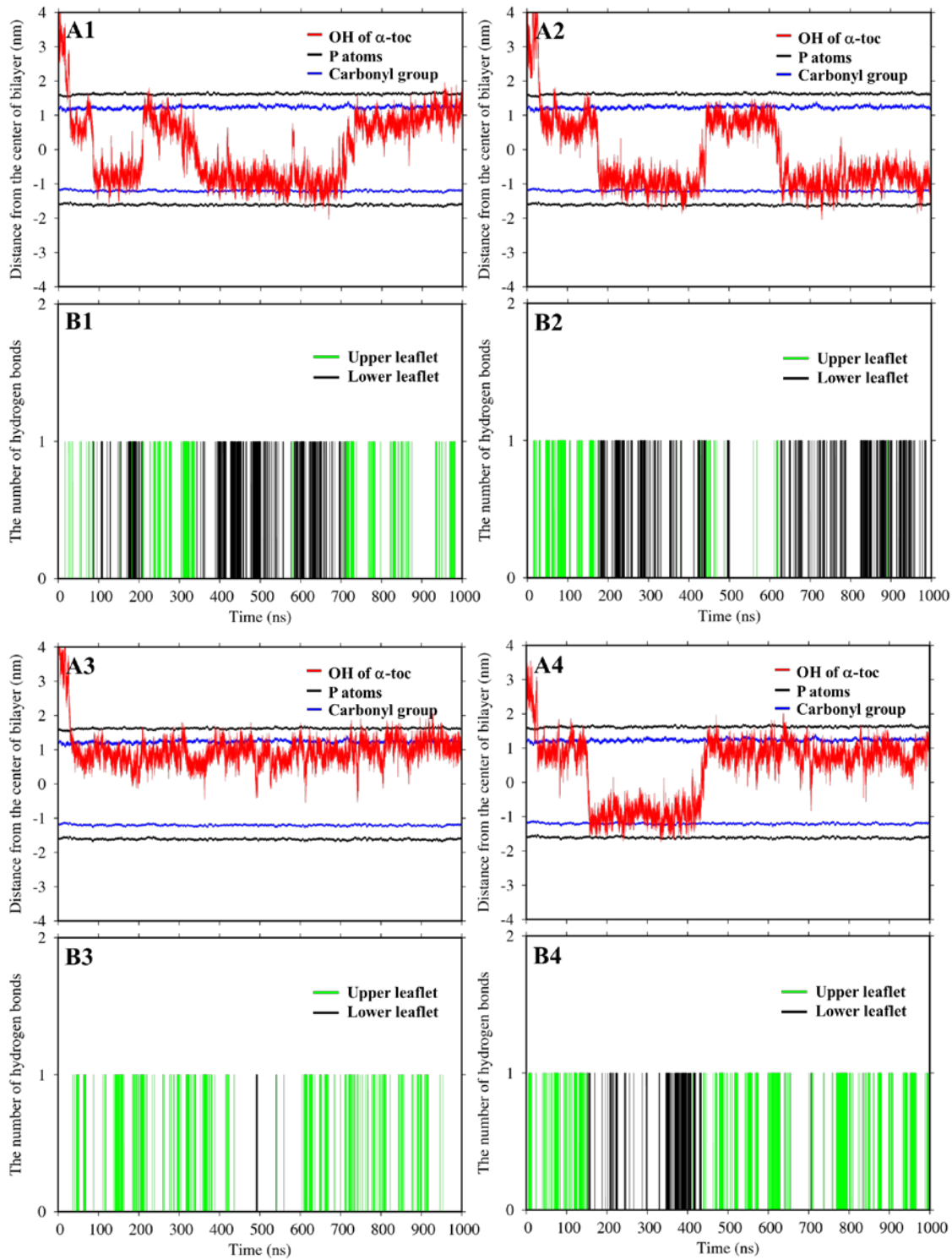

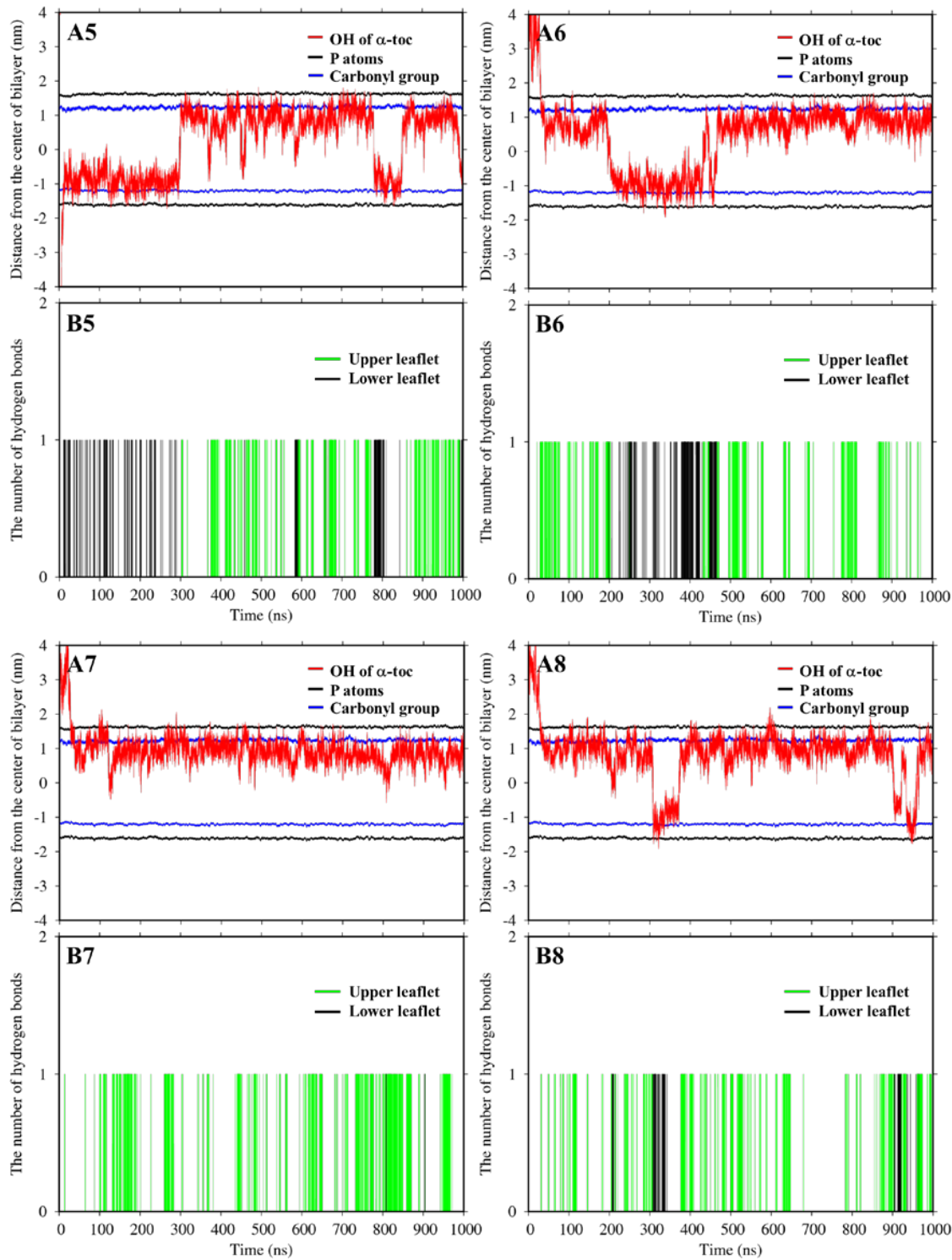

*Figure S1*. The time evolution of the position in along the z-direction for the hydroxyl group for each α-toc from the center of 50 % 9-al lipid bilayer with 8 α-toc molecules (A) and the number of hydrogen bonds between the hydroxyl group of α-toc and the aldehyde groups of the oxidized lipid tails in the upper and lower leaflets (B).